# Cross-Country Age Disparities in COVID-19 Cases with Hospitalization, ICU Usage, and Morbidity

Enes Eryarsoy [a] and Dursun Delen [b,#]

[c] Department of Management Information Systems, Sehir University,
Istanbul, Turkey. [eneseryarsoy@sehir.edu.tr]

[a] Department of Management Science, Oklahoma State University,
Stillwater, OK, USA. [dursun.delen@okstate.edu]

**Abstract**

In this paper, we examine cross-country differences, in terms of the age distribution of symptomatic cases, hospitalizations, intensive care unit (ICU) cases, and fatalities due to the novel COVID-19. By calculating conditional probabilities, we bridge country-level incidence data gathered from different countries and attribute the variability in data to country demographics. We then provide case, hospitalization, ICU, and fatality estimates for a comprehensive list of countries using the existing data from a variety of countries.

## 1. Introduction and Background

The first COVID-19 outbreak took place in the city of Wuhan in the Hubei province of China. Despite strict and robust prevention measures taken in the city, the virus has spread the rest of the world in a matter of a few weeks. Within three months, the World Health Organization (WHO) declared the outbreak a pandemic. While some more significantly than others, the virus has taken its toll on all countries with no exception. The global impact of COVID-19 has been very profound and probably unprecedented since the Spanish flu (H1N1 influenza circa 1918). Due to the novelty of the virus, and the nonexistence of vaccination, health professionals have been trying to cope with the pandemic using symptomatic treatment regimes. The rapid spread of the virus has caused a record influx of patients at hospitals, pushing the primary care health systems to the brink of total collapse. In order to ease the excessive burden on their healthcare systems, the governments are seeking out ways to suppress the transmission of the virus.

The rapid spread of the virus has even made the calculation of the rate of spread difficult. One frequently used way of measuring the spread is by computing the average number of secondary cases, or infections, that each case generates. This is known as the R-naught ($R_0$) of the virus. The $R_0$'s time and place dependent nature (typically smaller in the South Asian countries, depending on measures such as social distancing, isolation, partial or complete lockdowns, quarantines by the local authorities) is making modeling the spread of the virus a moving target (De Brouwer et al., 2020).

Even though the literature on COVID-19 is rapidly expanding, there is still a lack of consensus among academics and other scientists on the dynamics of the spread. This can perhaps be attributed to many reasons, such as the unpreparedness to a pandemic at this level, the lack of unified reporting systems due to diversity of health systems across the world, and the novelty of the pandemic itself. Many governments are seeking out forming different strategies that involve mitigating the spread until a method of prevention or a well-defined, and a successful treatment regime is found (Ferguson et al., 2020). The main focus of such mitigation effects is to alleviate the burden on healthcare systems by spreading out the diffusion of cases over a more extended period of time. While trying to achieve this, governments also face many uncertainties. One such uncertainty involves the absence of proven methods to accurately estimate the potential demand for healthcare services.

At the time of this paper's writing, several governments, such as Italy and Spain, already had over 100% health services capacity utilization, while others were about to experience a similar influx of critical patients. It is clear that governments are in need of better understanding the dynamics of the spread for optimal or near-optimal resource allocation decisions. Unfortunately, due to the emergency and the gravity of the pandemic and the lack of scantily found hard evidence cause such decisions to be made through the seat-of-the-pants approaches.

Perhaps one of the reasons behind the lack of evidence is that there is no obvious way to map reports and studies pertaining to one country into another. Many regional differences make this mapping and transfer the learnings and knowledge over to another domain particularly difficult. For the case of COVID-19, gender, and age of the patient populations seems to be among the key drivers of such differences. Academics are acting swiftly to enrich the medical literature by reporting their findings on the virus-related population characteristics, diffusion patterns, treatment regimes, case dynamics, hospitalizations, ICU usages, and fatalities.

In this paper, we build on the studies and reports that involve age-based clinical fatality risks (CFR), infection fatality risks (IFR), hospitalizations, ICU usages, and fatal outcomes. Using the latest literature as well as expert opinions, we attempt to combine data from different regions in order to estimate and highlight: (i) country-level differences, and (ii) healthcare system demands for individual age groups. Specifically, using the data from six different countries, we study the spread of the virus for different age groups.

While it is now known that the virus affects the elderly population more severely than the youngers, studies often report inconsistent results. Several underlying reasons may explain these inconsistencies. Perhaps one of the most plausible reason is the abundance of undocumented cases. In their study, (Li et al., 2020b) highlight that one of the reasons for the rapid spread of the virus is due to no documentation. They estimate that around 86% of all infections were undocumented. Another study from South Korea suggests similar undocumented case percentages at around 55-86% (Kim et al., 2020). News also suggests that even mortality cases often go unreported. A recent article in The Economist (Fatal flaws, 2020) highlighted stark differences between the number of expected death cases (including those attributed to COVID-19) and the actual death cases. Their estimation, based on regions' normal death rates, suggests the actual death-toll of the novel COVID-19 being more than double of what is being reported in different regions in Italy, Spain, and France. Perhaps this may be one of the reasons for conflicting CFR and IFR figures reported in the literature. While some studies suggested an estimated case fatality risk of as high as 7.2% (Onder et al., 2020) in Italy, other studies suggested a CFR of 3.4% in China (Wilson et al., 2020), 2.3% using the age-adjusted Diamond Princess cruise ship data (Russell et al., 2020). More recent works report somewhat lower case fatality rates circa 1.4-1.5%% (Guan et al., 2020; Wu et al., 2020), both using data from Wuhan.

Similar variations also hold for IFR. Studies report IFRs such as 0.5% using the Diamond Princess cruise ship data (Russell et al., 2020), 0.94%, and 0.657% using Wuhan data (Famulare, 2020) and (Verity et al., 2020), respectively. Even though these numbers significantly differ from each other, there seems to be (i) convergence to a CFR of 1.5% over time, (ii) CFR/IFR ratio appears to hover around 2-3, indicating as much as 40-70% asymptomatic cases of the virus.

One of the apparent reasons behind the differences in reported CFRs is the different demographics in different countries. As the virus affects the elderly more than the young, the virus takes a different toll on each country, depending on its age demographics. There may be several other reasons, including the fact that the age distribution of the infected population may differ from the overall age distribution. Hence, $R_0$ may significantly differ across age groups. For instance, a recent report indicated that the death-toll for under 60-year-old patients in Turkey is as much as four times higher than that of the death-rates in other European countries. As Turkey is one of the younger countries, this may be attributed to the demographical differences across countries or even mobility differences across the same age groups in different countries.

In this study, we believe that we make two main contributions to the existing literature. First, we investigate the role of age demographics on identified cases, hospitalization, ICU usages, and fatalities across different countries. And then, by applying conditional probabilities and the existing reports from different countries, we create a country independent conditional probability for each group. Second, by coupling expert-formed scenarios for the US, with age-standardized rates estimated from different countries, we calculate the age breakdown of symptomatic cases, hospitalizations, ICU usage, and fatalities.

In this study, we omit some critical analysis, such as taking into account gender differences. In essence, the virus seems to affect males more than females, due to scarcity of specific data, along with making some assumptions. For instance, we assume that deaths, hospitalizations, and ICU usages are proxy measures for COVID-19 spread. We also assume that similar spread patterns apply to each age group across countries—the virus is identical across all countries. We also demonstrate some evidence for the acceptability of these assumptions. The rest of this paper is organized as follows: in the next section, we describe our method and elaborate on the data used. The final section is dedicated to the findings and related discussions.

## 2. Method and Dataset

Many of the existing studies focus on modeling the spread of the COVID-19 have been using a few different models such as susceptible-infected-recovered (SIR) and its covariates (mainly SEIR: susceptible-exposed-infected-recovered) (Fang et al., 2020; Liu et al., 2020; Peng et al., 2020; Radulescu and Cavanagh, 2020), Sidarthe model (Giordano et al., 2020). These studies often ignore age-dependent variations from one country to another, or are limited to one country, or sometimes two (Ferguson et al., 2020).

In this study, instead of computing the spread of the virus, we look at the results of different scenarios. We base our analyses on expert opinions and the existing data on specific countries.

*Expert opinions:* Based on the characteristics of the COVID-19, Center for Disease Control (CDC) estimated that 2.4 to 21 million Americans would require hospitalization, and a death-toll of as much as 480,000 may be expected (Fink, 2020). According to the same projection, the death toll could be any figure from 200,000 to as high as 1.7 million. Another more recent estimate assuming full social distancing through May (as of April 8, 2020), the White House estimated this figure to fall between 30,000 and 126,000 (Institute for Health Metrics and Evaluation, 2020b).

While aggressive quarantines and enforcing/recommending social distancing can change the outcome of the burden on healthcare systems, the primary health care capacities are the bottleneck for almost all countries. The size of the susceptible population typically depends on different $R_0$ values. By enforcing/recommending social distancing, governments attempt to mitigate the situation and chance this figure (*Figure 1*).

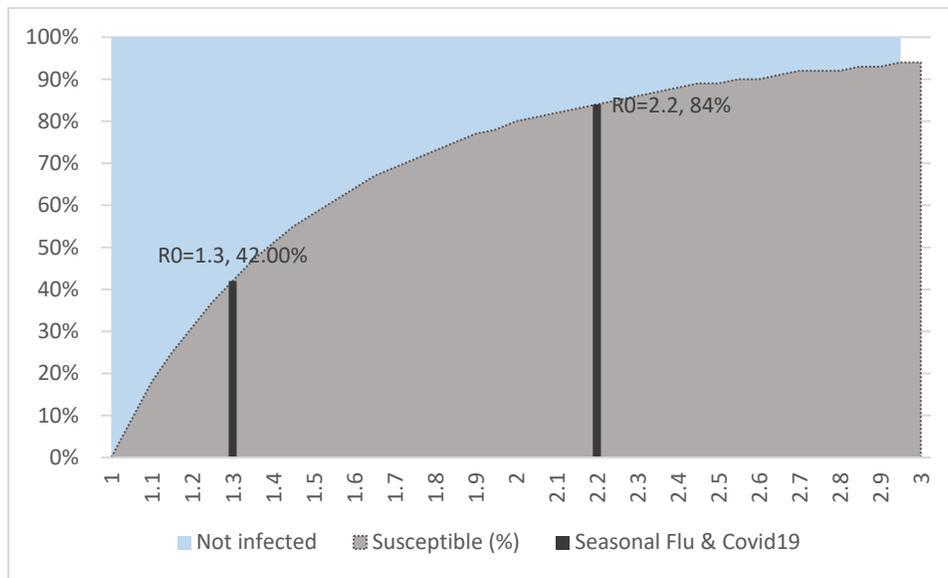

**Figure 1**. Different $R_0$ values and corresponding estimated percent of susceptible populations.

The novel coronavirus is often compared and contrasted against seasonal flu. Using CDC numbers over the last two flu seasons (2017-2019), we estimate the following table for the seasonal flu for comparison purposes.

**Table 1**. The seasonal flu numbers from CDC.

| Seasonal Flu (based on CDC data) | US Cases | As % of Susceptible Population | Per 1M |
|---|---|---|---|
| Susceptible population (R0=1.3) | 134.4M | 100.00% | 420,000 |
| Population with symptoms | 40M | 29.80% | 121K |
| Medical Visits | 15M | 11.10% | 45.5K |
| Hospitalization | 0.6M | 0.44% | 1,823 |
| Fatality | 40K | 0.03% | 121 |
| CFR | 0.03% | | |

We construct an analogous table to seasonal flu using expert opinions (Fink, 2020; Institute for Health Metrics and Evaluation, 2020b). After carefully scanning the existing literature, we built *Table 2* outlining our analysis. We constructed different scenarios from more severe (Scenario 1) to the least severe (Scenario 3) based on the estimates provided by the body of experts. We construct this table for the United States. By making use of expert expectations based on (Fink, 2020; Murray, 2020) we create a range of possible $R_0$ values to estimate the percentage of susceptible population. Using the literature, we then estimate upper and lower limits for symptomatic cases, reported cases (not all of the symptomatic cases are reported), hospitalizations (as a percentage of reported cases), as well as ICU cases (in terms of cases), and fatalities. One model (Murray, 2020) makes forecasts of the number of days will be needed, while Spain data can be used to calculate this number directly (Ministerio de Sanidad, 2020). After age-adjustments, we use data about hospitalizations as well as ICU usages from Spain. Assuming 10-15 days of average stay (as suggested in the literature), the data are consistent with those reported in (Murray, 2020).

Similarly, estimating fatalities also is difficult. Case fatality rates depend on a number of factors, including:

(i) The number of tests (and therefore the number of positive cases) conducted each individual country. Many countries—excluding countries such as Iceland, where a significant portion of the population was tested—conduct selective testing. This may involve a selection bias where only the people with severe enough symptoms may be tested.

(ii) The delay between the symptom onsets and the time of deaths.

(iii) The varying levels of adequacy/inadequacy of the healthcare systems.

(iv) The rates of smoking or the prevalence of chronical illnesses. We chose to use CFR of 1.5% for the United States for our analysis.

Table 2. Different COVID-19 spread estimates for the United States (numbers are in millions)

| COVID-19 | Scenario 1 | Scenario 2 | Scenario 3 | Expert Opinion 1 | Expert Opinion 2 |
|---|---|---|---|---|---|
| Susceptible population (overall R0={2.2[1], 1.8, 1.5[2]}) | 276.4 | 240.2 | 190.8 | 160-210M | - |
| The population with symptoms {.50[3], .35, .20[4]}(C) | 138 | 84 | 38 | - | - |
| Reported cases (as the ratio of symptomatic cases) {.11[5], .15, .20[6]} | 27.6 | 13 | 4.2 | - | - |
| Hospitalizations (36%[7,8] of reported cases)(H) | 10 | 4.5 | 1.5 | 2.4-21M | - |
| ICU patients (% of Hospitalizations–7.4%[9,10])(I) | 0.7 | 0.34 | 0.11 | - | ~0.08[11] |
| Fatalities (D) | 0.41 | 0.19 | 0.063 | 0.2-1.7M | 0.03-0.12 |
| CFR {1.5[12,13]} | 1.50% | 1.50% | 1.50% | | |

In this study, we use country age distributions (using population pyramids) and data involving different countries to create conditional probabilities. The severity of COVID-19 is also gender-dependent. However, due to the unavailability of data, we did not take gender into account.

---

[1] (Li et al., 2020a): estimates $R_0$ value as 2.2 for Wuhan data.
[2] (Shim et al., 2020) estimates $R_0$ value as 1.5 for South Korea where the virus is relatively better contained
[3] (Wu et al., 2020) finds a 50% chance of developing symptoms using Wuhan data.
[4] (Day, 2020) suggests that the ratio of asymptomatic cases could be as high as 80%.
[5] (Lachmann et al., 2020) uses South Korean data and similar age corrections and estimate that only around 11% of cases are reported in the US.
[6] (Magal and Webb, 2020) uses an estimation interval of 40-60% in their analysis. However, we construct our scenarios using similar to the US-based study.
[7] (CDCMMWR, 2020a) CDC estimates also yield a similar number in the US. However, the data has no age breakdown and includes some unknown cases.
[8] The rate is estimated as 0.36 using conditional probabilities and age distributions based on Spanish Data (Ministerio de Sanidad, 2020)
[9] (Murray, 2020) estimates that around a fifth of hospitalization days will require ICU stays for the US. However, they do not provide projections based on the number of hospitalization and ICU cases. Their numbers are in seem consistent with Spanish data (Ministerio de Sanidad, 2020).
[10] The crude rate is estimated as 7.4% of total hospitalizations using Spanish Data (Ministerio de Sanidad, 2020). However, each case stays in ICU for an average of 15 days (European Society of Anaesthesiology, 2020). (0.074 x 15 = 1.11)
[11] Assuming a 10-day average duration of stay.
[12] (Guan et al., 0; Wu et al., 2020) indicated a CFR of 1.5%.
[13] (Omer et al., 2020) indicates 0.7% CFR in Germany, such numbers usually come from countries where the spread is better contained.

We report our findings using the mildest of the three scenarios. We use the following notation:

> E: Events, E = {C: Case, A: Age, H: Hospitalization, I: Intensive Unit Care, D: Death)
>
> $P(C)$: The probability of being infected with symptoms. Using the scenario-2 with $R_0$=1.8, we estimate the proportion of the susceptible population as 0.73, with a 35% probability of developing symptoms: $P(C) = 0.35 \times 0.73$
>
> $P(R)$: The probability of being a reported case: $P(R) = P(C) \times .15$
>
> $P(H)$: The probability of hospitalization. This number depends on the percentage of reported cases, as well as the size of the population with symptoms. Using 0.15 and 36% of the rate of hospitalization we use $(P(H) = P(C) \times .11 \times .36)$
>
> $P(I)$: The probability of needing ICU $(P(I) = P(H) \times 0.074)$
>
> $P(D)$: The probability of death for the cases (CFR) $(P(D) = P(R) \times .015)$

$P(A_i)$: The probability of each age group *i* for a given country (using the country population pyramid)
$P(C|A_i)$: The probability of being infected with symptoms given age group *i*
$P(A_i|C)$: The probability of the age group *i*, given case.

We compute other conditional probabilities similarly for events {H, I, D}. We then use the conditional probabilities to simulate the mild scenario breakdowns for the United States. By using population pyramids and the US data, we replicate the same scenario for each individual country and report the results (per 1 million residents).

$$P(C|A) = \frac{P(A|C)}{P(A)} P(C)$$

A sample table, including some of the probabilities using the reports by the Spanish Ministry of Health, is given in *Table 3*. Using age-corrections via conditional probabilities also shows that reported numbers are quite consistent across-countries (*Table 4* and *Figure 4*).

Table 3. Probabilities and conditional probabilities for Spain

| Age Group | P(A) | #Conf. Cases | P(A|C) | #Hosp. | P(A|H) | #IUC cases | P(A|I) | #Deaths | P(A|D) |
|---|---|---|---|---|---|---|---|---|---|
| 0-9 | 9.3% | 130 | 0.6% | 35 | 0.45% | 1 | 0% | 0 | 0.0% |
| 10-19 | 10.0% | 226 | 1.1% | 20 | 0.26% | 1 | 0% | 1 | 0.1% |
| 20-29 | 10.0% | 1,352 | 6.6% | 200 | 2.6% | 10 | 2% | 4 | 0.5% |
| 30-39 | 13.2% | 2,386 | 11.7% | 431 | 5.6% | 18 | 3% | 3 | 0.4% |
| 40-49 | 17.0% | 3,190 | 15.6% | 778 | 10.1% | 45 | 8% | 9 | 1.1% |
| 50-59 | 14.9% | 3,433 | 16.8% | 1,074 | 13.9% | 106 | 18% | 20 | 2.5% |
| 60-69 | 11.1% | 3,179 | 15.6% | 1,432 | 18.6% | 162 | 28% | 63 | 7.8% |
| 70-79 | 8.4% | 3,304 | 16.1% | 1,858 | 24.1% | 192 | 34% | 164 | 20.4% |
| 80+ | 6.2% | 3,271 | 16.0% | 1,871 | 24.3% | 38 | 7% | 541 | 67.2% |

Table 4. Computing age group probabilities for given cases in (i) Spain, (ii) in the US calculated using Spain data[14] and (iii) reported by CDC[15]. While the correlation between (i) and (iii) is .88, the correlation between (i) and (ii) is as high as .97.

| Age Group | $P(A)_{US}$ | $P(A)_{Sp}$ | $P(A|C)_{Sp}$ | $P(A|C)_{US}$ from Spain data | $P(A|C)_{US}$ reported |
|---|---|---|---|---|---|
| 0-9 | 12.1% | 9.3% | 0.6% | 0.9% | 2.5% |
| 10-19 | 12.9% | 10.0% | 1.1% | 1.6% | 2.5% |
| 20-29 | 14.0% | 10.0% | 6.6% | 10.5% | 11.5% |
| 30-39 | 13.4% | 13.2% | 11.7% | 13.3% | 11.5% |
| 40-49 | 12.2% | 17.0% | 15.6% | 12.7% | 14.5% |
| 50-59 | 12.9% | 14.9% | 16.8% | 16.4% | 17.5% |
| 60-69 | 11.5% | 11.1% | 15.5% | 18.1% | 17.1% |
| 70-79 | 7.0% | 8.4% | 16.1% | 15.1% | 12.6% |
| 80+ | 3.9% | 6.2% | 16.0% | 11.4% | 10.2% |

We then use country demographics, CDC estimations for the US, and data sets available (Table) to compute age-adjusted probabilities and number of cases for each of the events (Susceptible, Case with symptoms, Hospitalization, IUC case, and Deaths) for countries with a population of more than 10 million people. We report all numbers per 1-million for ease of comparison.

---

[14] (Ministerio de Sanidad, 2020)
[15] (CDCMMWR, 2020b)

**Table 5.** Datasets used in this study

| Country | South Korea[1] | Spain[2] | US[3] | China[4] | Italy[5] |
|---|---|---|---|---|---|
| Number of cases | 6,284 | 20,471 | 2,449 | 44,669 | ~34,000 |
| Number of hospitalizations | - | 7,699 | - | - | - |
| Number of ICU cases | - | 573 | - | - | - |
| Number of fatalities | 42 | 805 | 44 | 805 | 1,625 |

[1].(Shim et al., 2020)
[2].(Ministerio de Sanidad, 2020)
[3].(Shim et al., 2020)
[4].(Verity et al., 2020)
[5].(Onder et al., 2020)

Studies report different case- and death-related age-breakdowns for a variety of countries. We observed that taking conditional probabilities—based on population age distributions in individual countries—into account, can help mitigate the variability in the reported results (*Figure 2*).

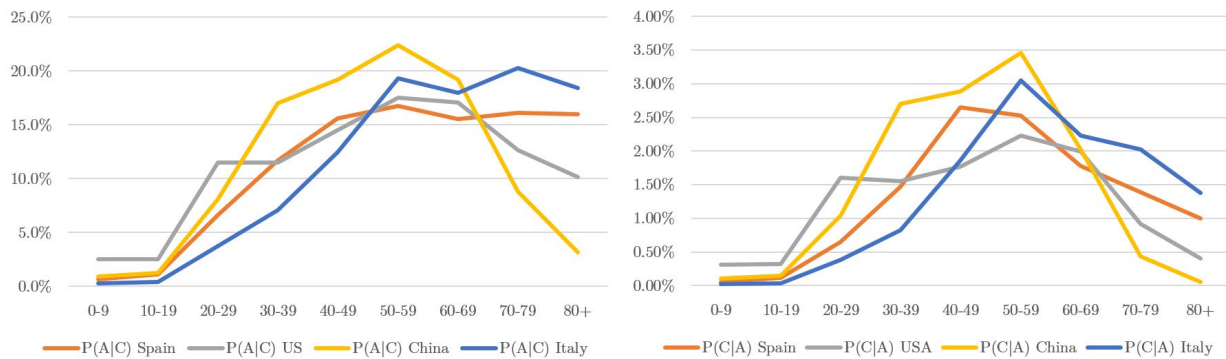

**Figure 2**. Age breakdown patterns of cases with (left) and without (right) taking country population pyramids (in terms of conditional probabilities) into account.

*Figure 3* also highlights the age differences for different events. ICU beds and invasive ventilators are in short supply, and some health systems prioritize younger patients over the older ones in order to increase the chances of survival. While debated, the figure also demonstrates such preferences.

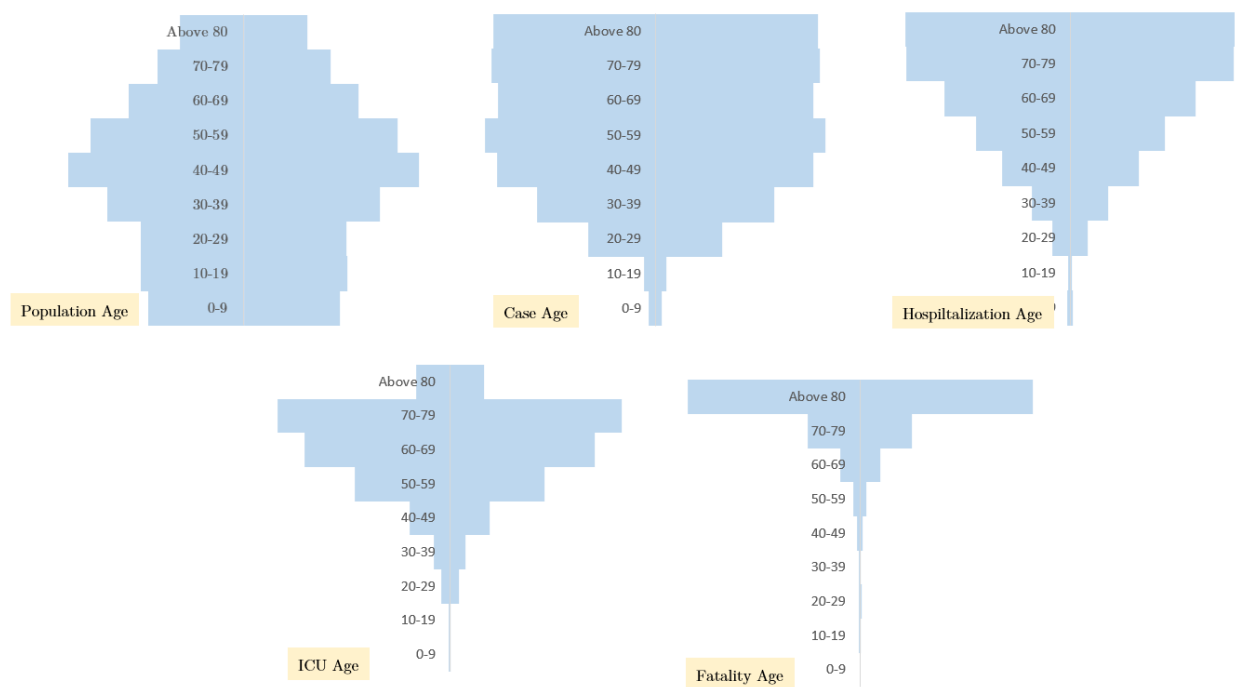

**Figure 3**. Age pyramids for cases, hospitalization, ICU uses, and fatalities for Spain.

Unfortunately, the only source of data that explicitly provided the age breakdowns of hospitalization and ICU cases (P(A|H), and P(A|I)) we could found was Spain (Ministerio de Sanidad, 2020). After comparing the estimated P(H) and P(I)s from Spain data and the data from the Institute for Health Metrics and Evaluation (2020), we concluded the numbers are consistent and decided to use age breakdowns from the Spanish dataset.

**Table 6**. Conditional probabilities for hospitalization and ICU using Spain data.

| Age Group | Hospitalized | P(A|H) | ICU | P(A|I) |
|---|---|---|---|---|
| 0-9 | 35 | 0.5% | 1 | 0.2% |
| 10-19 | 20 | 0.3% | 1 | 0.2% |
| 20-29 | 200 | 2.6% | 10 | 1.7% |
| 30-39 | 431 | 5.6% | 18 | 3.1% |
| 40-49 | 778 | 10.1% | 45 | 7.9% |
| 50-59 | 1,074 | 13.9% | 106 | 18.5% |
| 60-69 | 1,432 | 18.6% | 162 | 28.3% |
| 70-79 | 1,858 | 24.1% | 192 | 33.5% |
| 80+ | 1,871 | 24.3% | 38 | 6.6% |

## 3. Results and Discussion

In this paper, we focus on age-dependent breakdowns of cases, hospitalizations, ICU usages, and fatalities (events) using a range of scenarios. We construct these scenarios by using expert views and existing reports in the literature and based on the US data. We then use conditional probabilities to compute age-standardized breakdowns for the events for all individual countries.

Our results propose a few important implications. Firstly, the results highlight the effect of demographical differences across countries on COVID-19 spread. They suggest that, provided everything else remains the same, the death toll difference due to age demographics could be as much as 20 times (Niger vs. Japan, *Figure 4*).

Secondly, our results have the potential to help decision-makers to accommodate age-specific aspects of the spread. Creating different age-based isolation strategies, depending on the age-demographics of individual countries may be considered.

Also, our study attempts to combine several parameters calculated or taken from different academic papers, reports, or data sources together in creating a range of scenarios. While this approach provides a somewhat holistic view of the phenomenon, it also omits other country-level differences such as social isolation policies, prevention strategies, the effectiveness of the individual health-care systems. Our final table (Appendix-1) must be interpreted as a comparison tool for different countries' exposure to the virus.

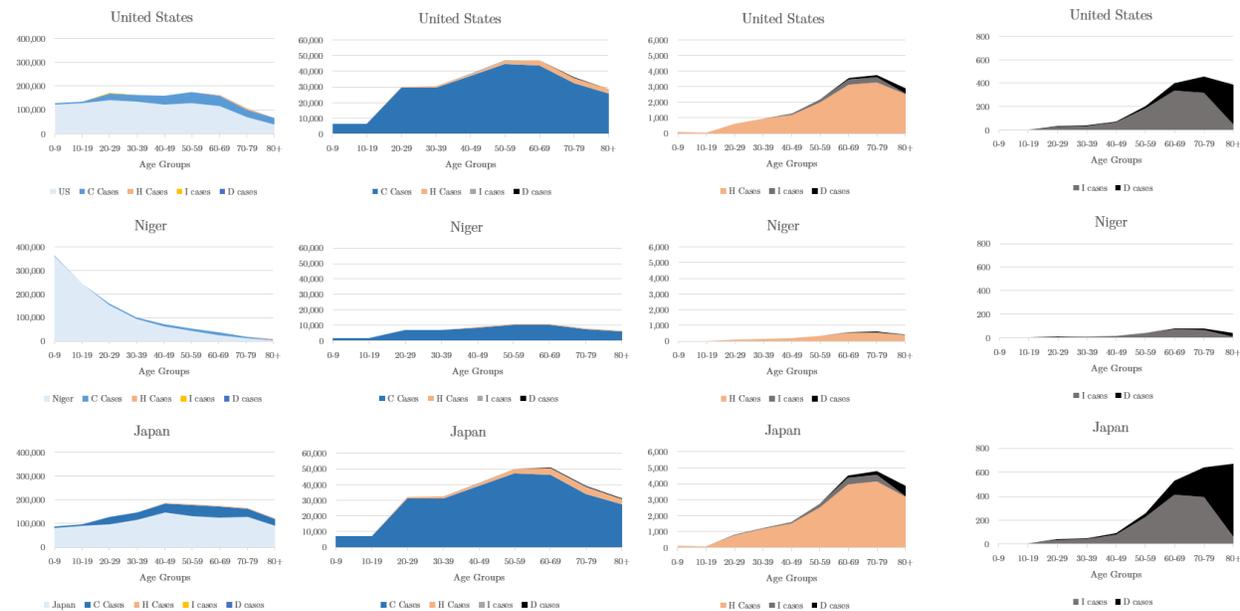

**Figure 4**. Age-dependent event estimations for the US, the country with the youngest population in the world (Niger), and with the oldest (Japan).

# References

CDCMMWR (2020a). Preliminary Estimates of the Prevalence of Selected Underlying Health Conditions Among Patients with Coronavirus Disease 2019 — United States, February 12–March 28, 2020. MMWR Morb. Mortal. Wkly. Rep. *69*.

CDCMMWR (2020b). Severe Outcomes Among Patients with Coronavirus Disease 2019 (COVID-19) — United States, February 12–March 16, 2020. MMWR Morb. Mortal. Wkly. Rep. *69*.

Day, M. (2020). COVID-19: four-fifths of cases are asymptomatic, China figures indicate (British Medical Journal Publishing Group).

De Brouwer, E., Raimondi, D., and Moreau, Y. (2020). Modeling the COVID-19 outbreaks and the effectiveness of the containment measures adopted across countries. MedRxiv.

European Society of Anaesthesiology (2020). Analysis of COVID-19 data on numbers in intensive care from Italy: European Society of Anaesthesiology (ESA).

Famulare, M. (2020). 2019-nCoV: Preliminary Estimates of the Confirmed-Case-Fatality-Ratio and Infection-Fatality-Ratio, and Initial Pandemic Risk Assessment. v2. 0.

Fang, Y., Nie, Y., and Penny, M. (2020). Transmission dynamics of the COVID-19 outbreak and effectiveness of government interventions: A data-driven analysis. J. Med. Virol.

Ferguson, N., Laydon, D., Nedjati Gilani, G., Imai, N., Ainslie, K., Baguelin, M., Bhatia, S., Boonyasiri, A., Cucunuba Perez, Z., Cuomo-Dannenburg, G., et al. (2020). Report 9: Impact of non-pharmaceutical interventions (NPIs) to reduce COVID19 mortality and healthcare demand.

Fink, S. (2020). Worst-Case Estimates for U.S. Coronavirus Deaths. N. Y. Times.

Giordano, G., Blanchini, F., Bruno, R., Colaneri, P., Di Filippo, A., Di Matteo, A., Colaneri, M., and others (2020). A SIDARTHE Model of COVID-19 Epidemic in Italy. ArXiv Prepr. ArXiv200309861.

Guan, W., Ni, Z., Hu, Y., Liang, W., Ou, C., He, J., Liu, L., Shan, H., Lei, C., Hui, D.S.C. (2020). Clinical Characteristics of Coronavirus Disease 2019 in China. N. Engl. J. Med. *0*, null.

Institute for Health Metrics and Evaluation (2020). IHME | COVID-19 Projections.

Kim, D.-H., Choe, Y.J., and Jeong, J.-Y. (2020). Understanding and Interpretation of Case Fatality Rate of Coronavirus Disease 2019. J Korean Med Sci *35*.

Lachmann, A., Jagodnik, K.M., Giorgi, F.M., and Ray, F. (2020). Correcting under-reported COVID-19 case numbers: estimating the true scale of the pandemic. MedRxiv.

Li, Q., Guan, X., Wu, P., Wang, X., Zhou, L., Tong, Y., Ren, R., Leung, K.S.M., Lau, E.H.Y., Wong, J.Y., et al. (2020a). Early Transmission Dynamics in Wuhan, China, of Novel Coronavirus–Infected Pneumonia. N. Engl. J. Med. *382*, 1199–1207.

Li, R., Pei, S., Chen, B., Song, Y., Zhang, T., Yang, W., and Shaman, J. (2020b). Substantial undocumented infection facilitates the rapid dissemination of novel coronavirus (SARS-CoV2). Science.

Liu, Y., Gayle, A.A., Wilder-Smith, A., and Rocklöv, J. (2020). The reproductive number of COVID-19 is higher compared to SARS coronavirus. J. Travel Med.

magal, pierre, and Webb, G. (2020). Predicting the number of reported and unreported cases for the COVID-19 epidemic in South Korea, Italy, France and Germany. MedRxiv.

Ministerio de Sanidad (2020). Actualización nº 54. Enfermedad por el coronavirus (COVID-19).

Murray, C.J. (2020). Forecasting COVID-19 impact on hospital bed-days, ICU-days, ventilator-days and deaths by US state in the next 4 months. MedRxiv.

Omer, S.B., Malani, P., and del Rio, C. (2020). The COVID-19 Pandemic in the US: A Clinical Update. JAMA.

Onder, G., Rezza, G., and Brusaferro, S. (2020). Case-fatality rate and characteristics of patients dying in relation to COVID-19 in Italy. JAMA.

Peng, L., Yang, W., Zhang, D., Zhuge, C., and Hong, L. (2020). Epidemic analysis of COVID-19 in China by dynamical modeling. ArXiv Prepr. ArXiv200206563.

Radulescu, A., and Cavanagh, K. (2020). Management strategies in a SEIR model of COVID 19 community spread.

Russell, T.W., Hellewell, J., Jarvis, C.I., van-Zandvoort, K., Abbott, S., Ratnayake, R., Flasche, S., Eggo, R.M., Kucharski, A.J., group, C. nCov working, et al. (2020). Estimating the infection and case fatality ratio for COVID-19 using age-adjusted data from the outbreak on the Diamond Princess cruise ship. MedRxiv.

Shim, E., Tariq, A., Choi, W., Lee, Y., and Chowell, G. (2020). Transmission potential and severity of COVID-19 in South Korea. Int. J. Infect. Dis. *93*, 339–344.

Verity, R., Okell, L.C., Dorigatti, I., Winskill, P., Whittaker, C., Imai, N., Cuomo-Dannenburg, G., Thompson, H., Walker, P.G., Fu, H., et al. (2020). Estimates of the severity of coronavirus disease 2019: a model-based analysis. Lancet Infect. Dis.

Wilson, N., Kvalsvig, A., Barnard, L.T., and Baker, M.G. (2020). Case-Fatality Risk Estimates for COVID-19 Calculated by Using a Lag Time for Fatality. Emerg. Infect. Dis. *26*.

Wu, J.T., Leung, K., Bushman, M., Kishore, N., Niehus, R., de Salazar, P.M., Cowling, B.J., Lipsitch, M., and Leung, G.M. (2020). Estimating clinical severity of COVID-19 from the transmission dynamics in Wuhan, China. Nat. Med.

Fatal flaws (2020). COVID-19's death toll appears higher than official figures suggest | Graphic detail | The Economist.

# Appendix 1

Table 7. Scenario-2 based event breakdowns for individual countries.

| Population (K) | Cases per 1M | Hosp. per 1M | ICU per 1M | Deaths per 1M | Country |
|---:|---:|---:|---:|---:|---|
| 10,196 | 275,770 | 16,717 | 1,218 | 854 | Portugal |
| 83,783 | 275,671 | 16,521 | 1,194 | 852 | Germany |
| 60,461 | 275,463 | 16,903 | 1,223 | 902 | Italy |
| 10,423 | 273,517 | 16,472 | 1,173 | 879 | Greece |
| 8,655 | 272,148 | 15,554 | 1,130 | 720 | Switzerland |
| 6,949 | 271,055 | 16,053 | 1,216 | 706 | Bulgaria |
| 126,475 | 271,042 | 17,430 | 1,256 | 1,037 | Japan |
| 9,660 | 270,439 | 15,464 | 1,160 | 665 | Hungary |
| 5,542 | 270,389 | 16,135 | 1,195 | 784 | Finland |
| 17,136 | 269,922 | 15,626 | 1,169 | 705 | Netherlands |
| 46,733 | 269,495 | 15,673 | 1,117 | 772 | Spain |
| 51,270 | 269,133 | 14,500 | 1,102 | 556 | Republic of Korea |
| 19,236 | 268,786 | 15,330 | 1,133 | 677 | Romania |
| 37,847 | 267,820 | 15,109 | 1,113 | 656 | Poland |
| 11,589 | 267,496 | 15,314 | 1,104 | 736 | Belgium |
| 5,792 | 266,784 | 15,340 | 1,144 | 694 | Denmark |
| 10,708 | 265,466 | 15,251 | 1,153 | 642 | Czechia |
| 10,099 | 264,504 | 15,248 | 1,108 | 729 | Sweden |
| 67,886 | 264,035 | 14,909 | 1,078 | 694 | United Kingdom |
| 43,734 | 263,634 | 14,445 | 1,065 | 598 | Ukraine |
| 5,421 | 261,559 | 14,405 | 1,060 | 619 | Norway |
| 11,326 | 260,576 | 13,965 | 1,038 | 562 | Cuba |
| 9,449 | 256,685 | 13,641 | 1,004 | 548 | Belarus |
| 329,064 | 255,500 | 13,797 | 1,021 | 575 | US |
| 4,822 | 254,069 | 13,778 | 1,022 | 576 | New Zealand |
| 25,498 | 253,440 | 13,696 | 993 | 590 | Australia |
| 145,935 | 253,217 | 13,567 | 997 | 551 | Russian Federation |
| 3,474 | 244,333 | 12,945 | 912 | 586 | Uruguay |
| 4,938 | 242,002 | 12,670 | 941 | 493 | Ireland |
| 19,116 | 234,534 | 11,579 | 853 | 429 | Chile |
| 1,439,324 | 227,809 | 10,805 | 875 | 326 | China |

| Population (K) | Cases per 1M | Hosp. per 1M | ICU per 1M | Deaths per 1M | Country |
|---|---|---|---|---|---|
| 45,197 | 213,004 | 10,468 | 759 | 400 | Argentina |
| 8,655 | 211,875 | 10,745 | 771 | 438 | Israel |
| 21,413 | 211,258 | 10,060 | 822 | 304 | Sri Lanka |
| 212,560 | 207,735 | 9,592 | 722 | 318 | Brazil |
| 50,883 | 198,163 | 9,060 | 679 | 301 | Colombia |
| **7,794,799** | **197,313** | **9,131** | **688** | **304** | **World** |
| 11,819 | 196,661 | 8,884 | 680 | 281 | Tunisia |
| 84,339 | 195,387 | 8,928 | 677 | 290 | Turkey |
| 32,972 | 191,468 | 8,717 | 649 | 290 | Peru |
| 97,338 | 190,509 | 8,391 | 608 | 284 | Viet Nam |
| 18,776 | 181,352 | 8,042 | 604 | 258 | Kazakhstan |
| 28,437 | 180,853 | 8,072 | 628 | 247 | Venezuela (Bolivarian Republic of) |
| 128,933 | 177,978 | 7,887 | 589 | 256 | Mexico |
| 10,847 | 174,998 | 7,815 | 574 | 263 | Dominica |
| 10,139 | 173,391 | 7,261 | 557 | 214 | Azerbaijan |
| 17,643 | 173,165 | 7,726 | 571 | 257 | Ecuador |
| 36,910 | 172,690 | 7,503 | 604 | 214 | Morocco |
| 32,365 | 168,392 | 7,211 | 575 | 205 | Malaysia |
| 11,674 | 163,184 | 7,436 | 533 | 267 | Bolivia |
| 83,993 | 161,603 | 6,787 | 541 | 189 | Iran (Islamic Republic of) |
| 43,852 | 159,997 | 6,934 | 521 | 217 | Algeria |
| 7,132 | 155,116 | 6,797 | 516 | 213 | Paraguay |
| 1,380,004 | 154,638 | 6,475 | 529 | 176 | India |
| 273,523 | 154,124 | 6,343 | 531 | 165 | Indonesia |
| 164,690 | 142,614 | 5,915 | 448 | 177 | Bangladesh |
| 109,581 | 136,985 | 5,625 | 457 | 152 | Philippines |
| 29,138 | 133,698 | 5,539 | 465 | 145 | Nepal |
| 59,308 | 132,704 | 5,360 | 455 | 135 | South Africa |
| 6,031 | 129,106 | 5,120 | 398 | 143 | Turkmenistan |
| 102,335 | 129,065 | 5,306 | 440 | 140 | Egypt |
| 33,470 | 128,528 | 5,038 | 401 | 135 | Uzbekistan |
| 17,500 | 125,473 | 5,083 | 400 | 141 | Syrian Arab Republic |
| 220,892 | 112,174 | 4,521 | 370 | 119 | Pakistan |

| Population (K) | Cases per 1M | Hosp. per 1M | ICU per 1M | Deaths per 1M | Country |
|---|---|---|---|---|---|
| 10,205 | 108,890 | 4,301 | 345 | 114 | Jordan |
| 34,815 | 102,129 | 3,777 | 317 | 91 | Saudi Arabia |
| 43,849 | 93,224 | 3,681 | 308 | 93 | Sudan |
| 114,964 | 88,752 | 3,545 | 291 | 93 | Ethiopia |
| 40,223 | 88,523 | 3,410 | 280 | 87 | Iraq |
| 40,223 | 88,523 | 3,410 | 280 | 87 | Iraq |
| 11,195 | 85,682 | 3,366 | 285 | 84 | South Sudan |
| 5,101 | 85,658 | 3,307 | 278 | 82 | State of Palestine |
| 27,692 | 82,226 | 3,138 | 265 | 77 | Madagascar |
| 31,073 | 79,587 | 2,934 | 280 | 63 | Ghana |
| 16,745 | 77,735 | 2,989 | 261 | 71 | Senegal |
| 89,561 | 76,412 | 2,970 | 255 | 72 | Democratic Republic of the Congo |
| 29,825 | 74,766 | 2,840 | 250 | 67 | Yemen |
| 31,255 | 71,548 | 2,752 | 239 | 66 | Mozambique |
| 15,893 | 70,564 | 2,714 | 242 | 63 | Somalia |
| 26,378 | 68,960 | 2,558 | 243 | 55 | Côte d'Ivoire |
| 26,545 | 68,088 | 2,540 | 232 | 57 | Cameroon |
| 59,734 | 66,731 | 2,484 | 229 | 55 | United Republic of Tanzania |
| 38,928 | 66,159 | 2,469 | 227 | 55 | Afghanistan |
| 53,771 | 66,064 | 2,407 | 219 | 53 | Kenya |
| 206,139 | 60,712 | 2,193 | 228 | 43 | Nigeria |
| 20,903 | 60,006 | 2,200 | 209 | 47 | Burkina Faso |
| 20,249 | 59,879 | 2,230 | 207 | 49 | Mali |
| 24,207 | 59,262 | 2,215 | 215 | 47 | Niger |